\documentclass[twocolumn,showpacs,amsfonts]{revtex4}
\usepackage{amsmath}
\usepackage{latexsym}
\usepackage{float}
\usepackage{amssymb}
\usepackage{graphicx}
\usepackage{textcomp}
\usepackage{hyperref}

\def\ba{\begin{eqnarray}}
\def\ea{\end{eqnarray}}
\def\be{\begin{equation}}
\def\ee{\end{equation}}

\def\bm{\begin{math}}
\def\me{\end{math}}

\newcommand{\dummy}

\begin{document}

\title {Diffusive Domain Coarsening : Early Time Dynamics and Finite-Size Effects}
\author{Suman Majumder and Subir K. Das$^{*}$}
\affiliation{Theoretical Sciences Unit, Jawaharlal Nehru Centre 
for Advanced Scientific Research, Jakkur P.O, Bangalore 560064, India}

\date{\today}
\begin{abstract}
~We study diffusive dynamics of phase separation in a binary mixture, following 
critical quench, both in spatial dimensions $d=2$ and $d=3$. Particular 
focus in this work is to obtain information about effects of system size 
and correction to the growth law via appropriate application of finite-size scaling
method to the results obtained from Kawasaki exchange Monte Carlo simulation of Ising model.
 Observations of only weak size effects and 
very small correction to scaling in the growth law are significant. The methods used in this work and information 
thus gathered will be of paramount importance in the study of kinetics of 
phase separation in fluids and other problems of growing length scale. 
We also provide detailed discussion on standard methods of understanding simulation 
results which may lead to inappropriate conclusions.

\end{abstract}
\maketitle
\section{INTRODUCTION}\label{intro}
~When a homogeneous binary mixture (A+B) is quenched inside the miscibility
gap, the system falls out of equilibrium and moves towards its new equilibrium 
state via formation and growth of domains rich in A- or B- particles 
\cite{Bray,Haasen,Wadhawan}. This coarsening of domains is a scaling phenomenon,
e.g., two-point equal-time correlation function $C(r,t)$, the structure factor 
$S(k,t)$ and the domain size distribution function $P(\ell_{d},t)$ obey the scaling 
relations
\begin{eqnarray}\label{scaledCr}
C(r,t) &\equiv& \tilde {C}(r/\ell(t)),\\
S(k,t) &\equiv&\ell^{d} \tilde{S}(k\ell(t)),\\
P(\ell_{d},t) &\equiv& \ell(t)^{-1}\tilde {P}[\ell_{d}/\ell(t)],
\end{eqnarray}
where the average domain size $\ell(t)$ increases with time ($t$) in a power-law fashion
\begin{eqnarray}\label{powerlaw}
 \ell(t)\sim t^{\alpha},
\end{eqnarray}
and $\tilde {C}(x)$, $\tilde {S}(y)$ and $\tilde {P}(z)$ are scaling functions 
independent of $\ell(t)$. In Eq. (\ref{powerlaw}), the growth exponent $\alpha$
depends upon the transport mechanism.
\par
~For diffusive growth, associating the rate of increase of $\ell(t)$ with the chemical
potential ($\mu$) gradient, one can write \cite{Bray} 

\begin{eqnarray}\label{dldt}
 \frac{d\ell(t)}{dt} \sim \lvert  \overrightarrow{\nabla} 
\mu \lvert \sim \frac{\sigma}{\ell(t)^{2}},
\end{eqnarray}
$\sigma$ being the A-B interfacial tension. Solution of Eq. (\ref{dldt}) gives 
$\alpha=1/3$, known as Lifshitz-Sloyozov (LS) law \cite{LSlaw}. The LS behavior 
is the only growth law expected for phase-separating solid mixtures. However,
 for fluids and polymers, one expects faster growth at large length scales where 
hydrodynamic effects are dominant. For the latter, in $d=3$, convective transport
yields additional growth regimes \cite{Siggia,Furukawa} with
\begin{eqnarray}\label{fluid_growth}
 \alpha&=&1,~~~~~\ell(t) \ll \ell_{in},\nonumber\\
&=&2/3,~~\ell(t) \gg \ell_{in}.
\end{eqnarray}
In Eq. (\ref{fluid_growth}), the inertial length $\ell_{in}$[$\simeq \eta/(\rho\sigma)
,\eta$ and $\rho$ being the shear viscosity and mass density] marks the crossover from 
a low-Reynold-number viscous hydrodynamic regime to an inertial regime.
\par
~Diffusive domain coarsening in solid binary mixtures has been extensively studied via 
Ising model
\begin{eqnarray}\label{Ising}
 H = -J\sum_{<ij>}S_i S_j;~S_{i}=\pm1,~J>0,
\end{eqnarray}
prototype for a large class of critical phase transitions. Here one can identify the spin 
$S_{i}=+1(-1)$ at lattice site $i$ with an A-particle (B-particle). Note that $<ij>$ 
in Eq. (\ref{Ising}) stands for summation over only the nearest neighbors. One can also
study the kinetics of phase separation via dynamical equations which can be obtained from
Ising models in mean field approximation by using a master equation
 approach \cite{Binder,Frisch} with Kawasaki exchange kinetics \cite{Kawasaki}. 
Upon coarse-graining, such equations lead to the Cahn-Hilliard (CH) equation
\begin{eqnarray}\label{Cahn-Hilliard}
 \frac{d\psi(\vec{r},t)}{dt}=-\bigtriangledown^{2}[\psi(\vec{r},t)+
\bigtriangledown^{2}\psi(\vec{r},t)-\psi^{3}(\vec{r},t)],
\end{eqnarray}
where $\psi(\vec{r},t)$ is a coarse-grained time-dependent local order parameter.
Note that such continuum description could also be obtained in a phenomenological manner 
\cite{Bray,Jone} using a coarse-grained Ginzburg-Landau(GL) free energy functional with 
the requirement of conservation of material. The CH equation with an 
added thermal noise is expected to be equivalent to Monte Carlo (MC) simulations
\cite{MCbook,Frankel} of kinetic Ising models.
\par
~In Eq. (\ref{Cahn-Hilliard}), typical
distance over which the order parameter is coarse-grained is of the size of equilibrium 
correlation length, $\xi$. In situations when one is interested in studying the kinetics 
in the close vicinity of the critical point, without focusing on the dynamics at 
microscopic level, Eq. (\ref{Cahn-Hilliard}) is computationally very useful in 
achieving the asymptotic \cite{Desai} behavior. However, for deep quenches 
one needs to incorporate higher order terms than are usually used in the GL hamiltonian. Also at very low temperature,
where $\xi$ is of the order of a lattice constant, CH equation would not 
provide information of a large effective system size compared to the atomistic Ising 
model. Particular focus of this work is to learn finite-size effects and dynamics at the early stage
both of which have received much less attention as opposed to the identification of 
domain growth law in long-time limit, despite their obvious importance both fundamentally as well 
as technologically, e.g, in nano-science and technology. In view of that, we choose to revisit 
kinetics of phase separation in Ising model via MC simulations.
\par
~While MC simulations have been used immensely in the understanding
of non-equilibrium domain growth phenomena both with conserved 
\cite{Barkema,Huse,Amar,Roland,Grant,Potts,Heermann,Vinals} and non-conserved 
\cite{Vinals,Milchev,Roland89}
order parameter, earlier studies of phase ordering in conserved Ising model with 
critical (50:50) composition reported \cite{MRao,Grest}
 estimates of $\alpha~\epsilon~[0.17,0.25]$, deviating drastically from the expected LS law.
Even arguments in favor of logarithmic growth were proposed \cite{Mazenko}. Note that 
these earlier reports were based on MC simulations
for very short period of time where contamination of domain structures due to thermal 
noise might not have been taken care of, which could act as a source of significant error in the
measurement of average domain size.
\par
~Later, the discrepancy of the previous results with the expected LS behavior was understood 
to be due to strong corrections to scaling at early time. To account for this \cite{Huse} 
higher order terms in Eq. (\ref{dldt}) were incorporated to write
\begin{eqnarray}\label{Correction}
\frac{d\ell(t)}{dt}=\frac{C_{1}}{\ell(t)^{2}}+
\frac{C_{2}}{\ell(t)^{3}}+O(\ell(t)^{-4}),
 \end{eqnarray}
which in the long time limit gives a solution $\propto t^{1/3}$, however, would give rise to leading order 
correction linear in $1/\ell(t)$ to the instantaneous exponent.
Thus LS behavior will be observed only in the limit 
$\ell(t\rightarrow \infty) \rightarrow \infty$.
Indeed, consistency with a linear correction was observed for 50:50
binary mixture \cite{Huse,Amar} as well as for multicomponent mixtures \cite{Potts}.
Present work, however, convincingly demonstrates that the observation of LS value of the 
exponent only in the asymptotic limit was misleading and presence 
of a time independent bare length in $\ell(t)$ was responsible for the numerical data 
exhibiting such trend.
\par
~Most of the studies till date, 
stressed on using large systems, with the anticipation of strong finite-size 
effects \cite{Shinozaki} combined with the expectation that the LS law will
be realized only in the large $\ell(t)$ limit. This strategy, of course, will prove to be 
useful when there is dynamical crossover as in domain coarsening in fluids 
[cf. Eq. (\ref{fluid_growth})] where the system size should be significantly 
larger than smallest characteristic length scale in a particular regime. 
However, consideration of arbitrarily large system sizes restricts the access of 
large time scale, particularly for 
molecular dynamics simulation of fluid phase separation \cite{Laradji,Thakre,Kabrede}.
 It is worth mentioning that the typical system sizes authors 
consider these days contain number of lattice sites or particles of the order of million, 
which is too large, even for present day computers, to access long time scale that often is a
necessity. Such choice of large systems, in addition to the anticipation of strong 
finite-size effect, was often motivated by the expectation of better self-averaging \cite{Shinozaki}.
However, our experience contradicts the latter and is rather consistent 
with the previous works \cite{Milchev} reporting lack of self averaging.
Thus a judicial choice of system sizes is very crucial for such problems \cite{Shaista} 
which in turn requires appropriate knowledge of finite-size effects \cite{Suman}. 
 While recent focus has been in more complicated systems 
\cite{Horbach,Mitchell,Bucior,Yelash,Yelash1,Das_Horbach,Hore,Das_Stat}
with realistic interactions and physical boundary conditions, many such basic information
 as discussed above are lacking even in a situation as simple as Ising systems.
\par
~In this work we have undertaken a comprehensive study to learn about the finite-size effects in domain
coarsening in Ising model with conserved order parameter dynamics and understand 
the behavior of growth exponent as a function of time, via application of finite-size scaling method 
\cite{Fisher,Privman}, both in space dimensions $d=2$ and $d=3$. Despite its glorious 
popularity in equilibrium critical phenomena, finite-size scaling method has been only rarely 
\cite{Heermann,Vinals,Suman} used in non-equilibrium processes. In this paper, we exploit this
 method appropriately in the context of diffusive phase separation kinetics to show that for 
critical quench the LS value of $\alpha$ sets in very early and the effect of size 
is very small.
\par
~This paper is organized as follows. In Sec.\ref{methods}, we describe the details of simulation 
and finite-size scaling method. Results for both $d=2$ and $d=3$ are presented in Sec. \ref{results},
 while Sec. \ref{summ} summarizes the paper with a discussion of future possibilities in this direction.
\section{Methods}\label{methods}
\subsection{Details of Simulation and Calculation of observables}\label{detail}
~In the MC simulation of Ising model, the conserved order-parameter dynamics, 
where composition of up (A particle) and down (B particle) spins remains fixed during 
the entire evolution, is implemented via standard Kawasaki exchange mechanism 
\cite{Kawasaki} where interchange of positions between a randomly chosen 
pair of nearest neighbor (nn) spins consists a trial move. A move is accepted or 
rejected according to standard Metropolis algorithm \cite{MCbook}. One MC step (MCS) consists 
of exchange trials over $L^{d}$ pairs of spins. Periodic boundary conditions were applied
in all directions. 
\begin{figure}[htb]
\centering
\includegraphics*[width=0.375\textwidth]{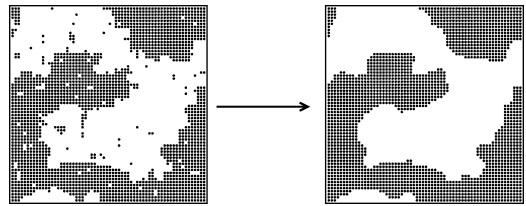}
\caption{\label{fig1} Left panel: Snapshot of a $2-d$ Ising model at $T=0.85T_{c}$
obtained from the Monte Carlo simulation via Kawasaki exchange kinetics, for $L=64$ at
$t=5\times10^{3}$ MCS. Right panel: Same snapshot after removing the noise via the 
exercise described in the text. A-particles are marked by black dots 
whereas B-particles are unmarked.}
\end{figure}
\par
~Note that with the increase of temperature,  accurate measurement of average domain size becomes
difficult due to the presence of noisy clusters of the size of $\xi(T)$. 
On the other hand, at very low temperature growth is hampered by 
metastability. To avoid the latter problem, 
we have set the temperature towards the higher side and calculated
 all the physical quantities from pure domain morphology after eliminating the thermal noise
via a majority spin rule. There a spin at a lattice site $i$ was replaced by the sign of the majority 
of the spins sitting at $i$ and nn of $i$ (depending upon the noise level i.e., average size of noise
clusters, extension to distant neighbors may also become necessary). In Fig. \ref{fig1} we demonstrate such filtering process
for a rather high temperature. The left panel corresponds to the original snapshot from the 
MC simulation on $2-d$ square lattice at $T=0.85T_{c}$ with $L=64$ at $t=5\times10^{3}$ MCS.
One can appreciate that presence of substantial noise elements here would give rise to smaller value of $\ell(t)$
than the actual. The right panel of the figure shows the snapshot with pure domain
morphology obtained after implementing the noise removal exercise described
above. Of course, one should be careful that too many such iterations 
or consideration of very distant neighbors may alter the basic structure. 
However, in the present case, no such deformation has taken place. All the quantities
in our simulation were calculated by using snapshots with such pure domain 
structure.
\begin{figure}[htb]
\centering
\includegraphics*[width=0.32\textwidth]{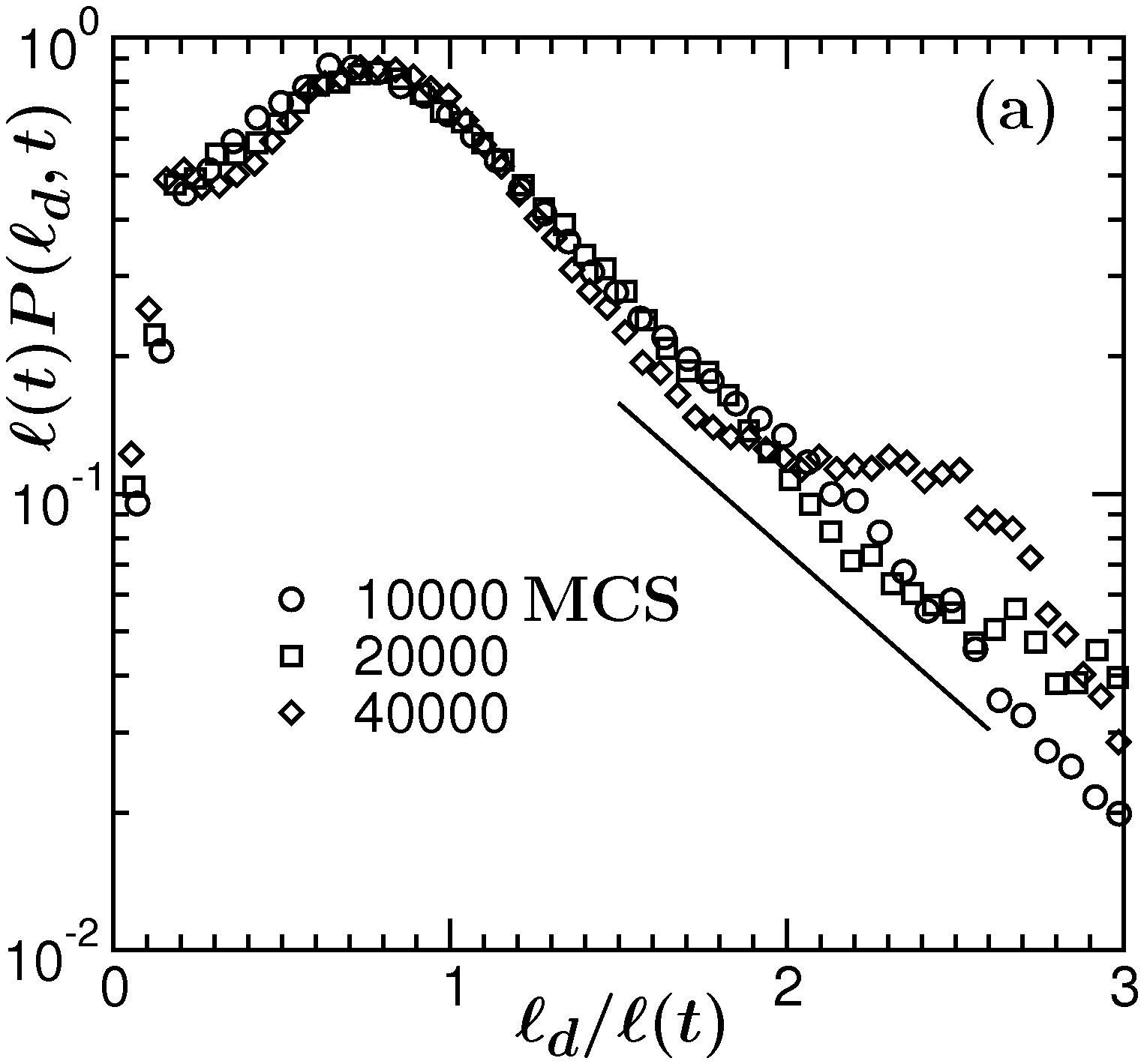}\\
\vskip 0.2cm
\centering
\includegraphics*[width=0.32\textwidth]{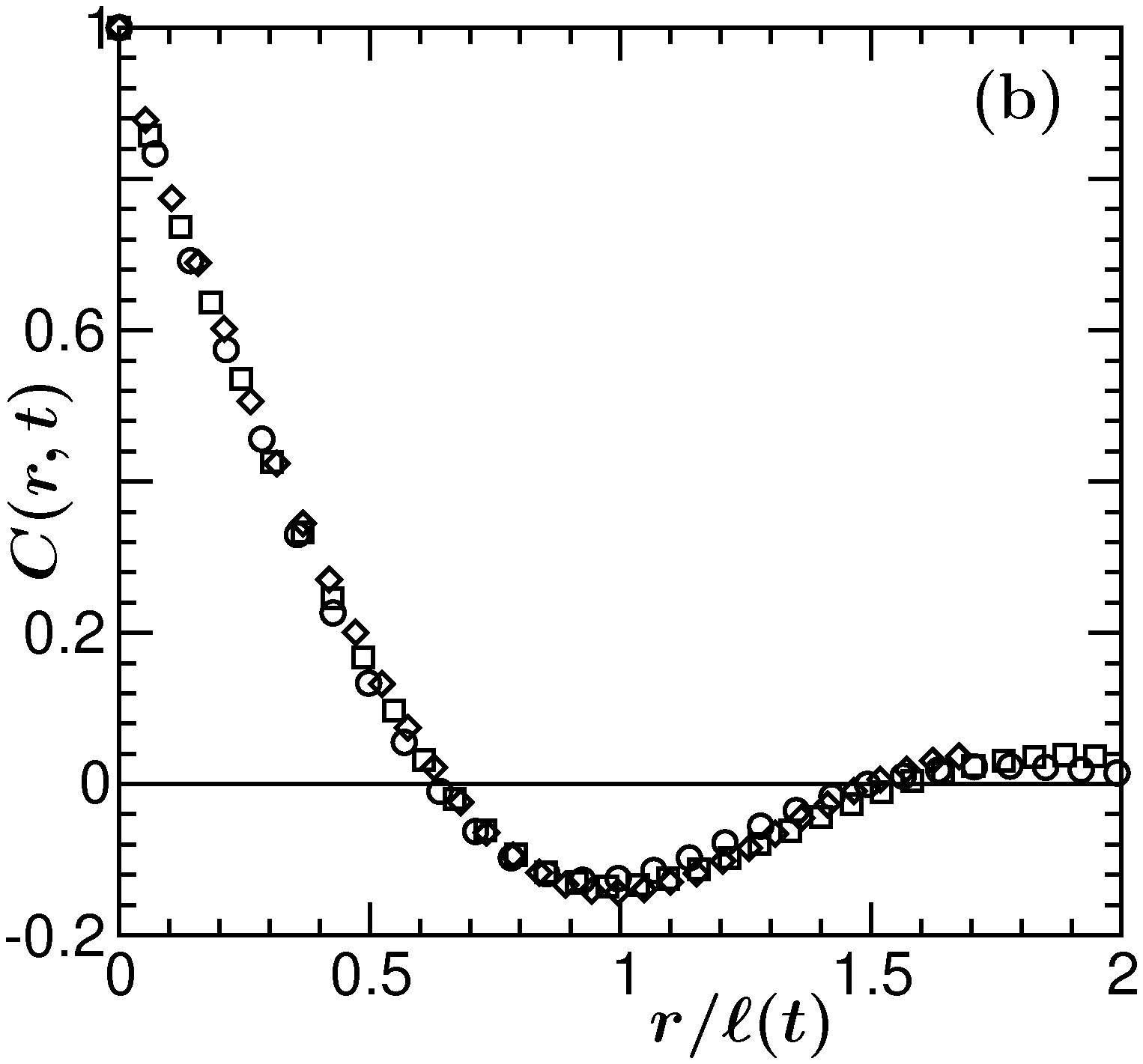}\\
\vskip 0.2cm
\centering
\includegraphics*[width=0.32\textwidth]{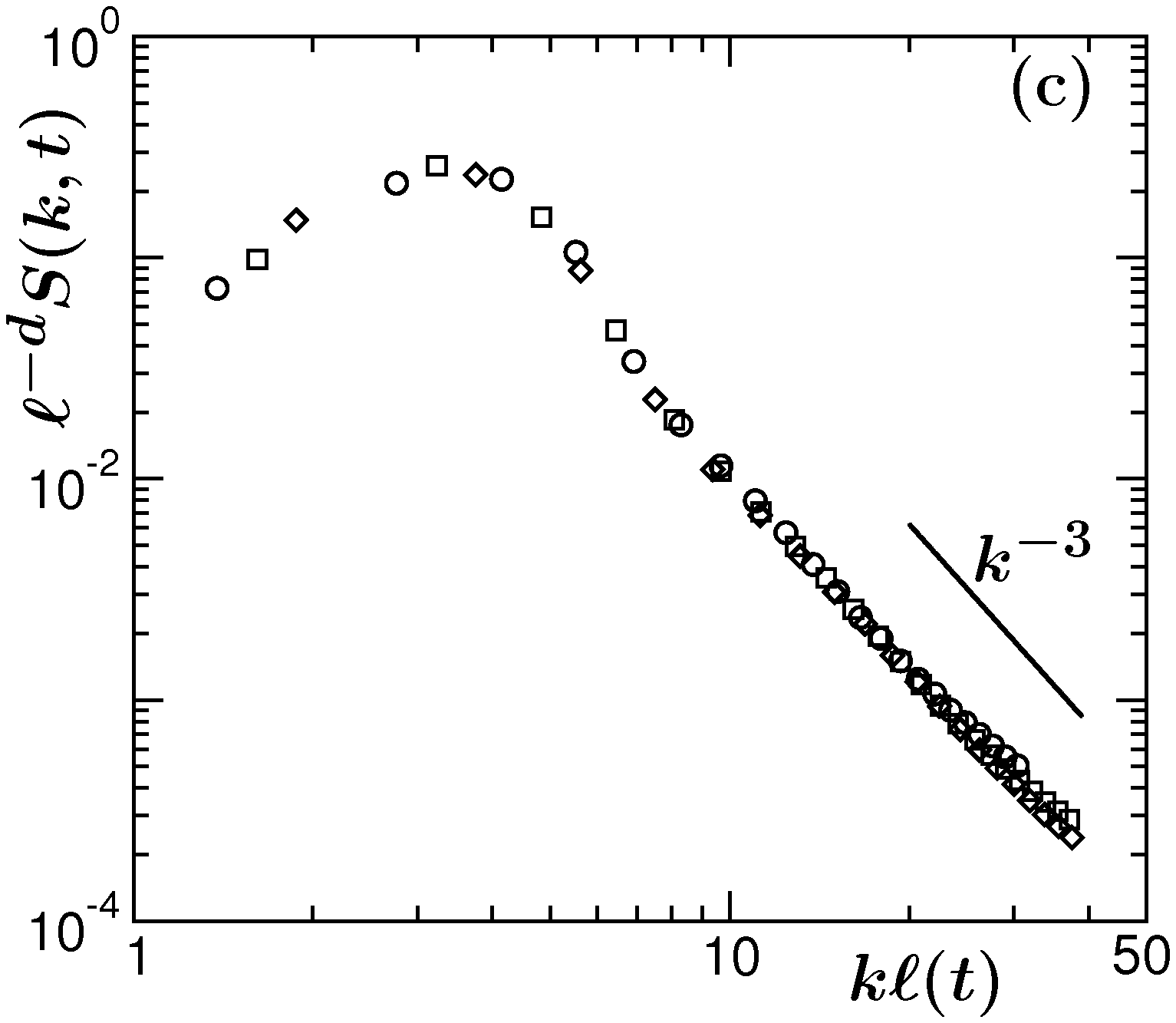}\\
\caption{\label{fig2} Scaling plots of (a) domain size distribution $P(\ell_{d},t)$, 
(b) correlation function $C(r,t)$ and (c) structure factor $S(k,t)$, from 
different times, as indicated, for the system in Fig. \ref{fig1}. The data were averaged 
over $50$ independent initial configurations. Note that in all the cases, $\ell(t)$ used, was calculated 
using Eq. (\ref{L_d}). While the solid line in (a) verifies exponential decay of the tail, 
the one in (c) corresponds to the Porod tail.}
\end{figure}
\par
~In Fig. \ref{fig2}(a) we present the scaling plots of domain size distribution function,
viz., plots of $\ell(t)P(\ell_{d},t)$ vs $\ell_{d}/\ell(t)$ where $\ell(t)$ was calculated from the
 first moment of the normalized distribution $P(\ell_{d},t)$ as
\begin{eqnarray}\label{L_d}
 \ell(t)=\int d\ell_{d}~\ell_{d} P(\ell_{d},t),
\end{eqnarray}
with length $\ell_{d}$ being the separation between two successive interfaces in $x$-,
$y$- or $z$- directions. Figs. \ref{fig2}(b) and \ref{fig2}(c) show 
the scaling plots of correlation function $C(r,t)$ and its Fourier transform $S(k,t)$, 
in accordance with Eq. (\ref{scaledCr}) and (2),
where $C(r,t)$ was calculated as
\begin{eqnarray}\label{Crt}
 C(r,t)=<S_{i}S_{j}>;~~r=|\vec{i}-\vec{j}|.
\end{eqnarray}
Note that these scaling plots for all the quantities were obtained by using the values of $\ell(t)$
obtained from Eq. (\ref{L_d}). Of course, independently $\ell(t)$ could be calculated 
from the decay of $C(r,t)$ as well the first moment of $S(k,t)$ as 
\begin{eqnarray}\label{lt_Crt}
 C(r=\ell(t),t)=h,
\end{eqnarray}
and
\begin{eqnarray}\label{lt_Skt}
\ell(t)=\frac{1}{\int dk~kS(k)}.
\end{eqnarray}
\par
~When calculated from a completely noise-free morphology, all the above mentioned methods
for the calculation of $\ell(t)$ must give results proportional to each other. 
When $h$ is set to a rather small value, particularly when the decay length is 
larger than the average size of the noisy clusters, calculation of $\ell(t)$ from 
Eq. (\ref{lt_Crt}) is not expected to be affected much by the presence of 
noise. The same applies for Eq. (\ref{lt_Skt}). However, when calculated via Eq. (\ref{L_d}), 
either the distribution up-to the length of average noise size should be appropriately modified 
or noise should be completely eliminated. The latter strategy is more appropriate since 
it gives better shape to the all form functions.
In our calculation, in Eq. (\ref{lt_Crt}), $h$ will correspond to first zero of $C(r,t)$. 
\par
~All the results presented in Fig. \ref{fig2} are obtained from pure domain morphology and the 
nice data collapse obtained in each case using the measure of $\ell(t)$ from Eq. (\ref{L_d}) 
speaks for the equivalence of all the definitions, Eqs. (\ref{L_d}), (\ref{lt_Crt}) and (\ref{lt_Skt}).
The linear behavior of the tail region in (a) on a semi-log plot is consistent with  
an exponential decay of $P(\ell_{d},t)$. Here the noisy look (oscillatory behavior) at late time or large domain 
size limit (which was also observed in other recent studies \cite{Sarrazin,Sicilia}) 
is due to lack of statistics when $\ell(t)$ becomes of the order of the system size.
We refer to it as a secondary finite-size effect which does not affect the growth dynamics severely.
On the other hand, the linear look of large wave vector ($k$) data in (c) confirms the Porod law 
\begin{eqnarray}\label{porod_law}
 S(k,t)\sim k^{-(d+n)}.
\end{eqnarray}
Note that in the present case $d=2$ and $n=1$ (symmetry of the order parameter). It is worth 
mentioning that one would have observed a slower decay of the structure factor had the 
noise not been removed.
\begin{figure}[htb]
\centering
\includegraphics*[width=0.35\textwidth]{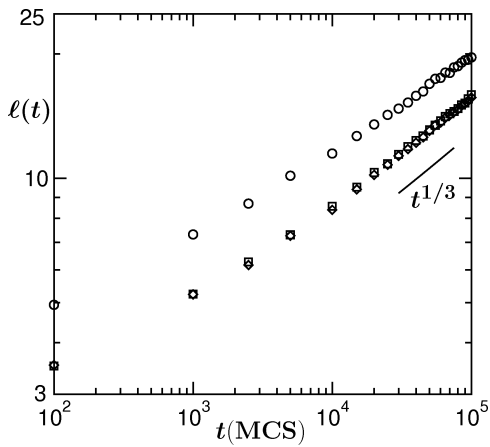}
\caption{\label{fig3} Average domain size is plotted on a log-scale as a function of 
time $t$. Different symbols correspond to calculation of $\ell(t)$ 
from different quantities : circles from $P(\ell_{d},t)$, squares from first 
zero-crossing of $C(r,t)$, diamonds from the first moment of $S(k,t)$. 
Results presented were obtained from pure domain structure as demonstrated in the 
right panel of Fig. \ref{fig1}, with $L^{2}=128^{2}$ and $T=0.85T_{c}$. 
The solid line corresponds to the theoretically expected $t^{1/3}$ behavior.}
\end{figure}
\par
~In Fig. \ref{fig3} we present the length scale results obtained from all the above mentioned methods 
on log scale, all of which look proportional to each other as was also clear from
the exercise of Fig. \ref{fig2}. The data at late times look consistent with the expected exponent 
$\alpha=1/3$. Note that if the temperature is sufficiently close to $T_{c}$, for long enough time 
the noise might not have equilibrated to the equilibrium value inside the true domains. 
In such a situation, there are two growing length scales in the problem which may give rise
to misleading value of the exponent if the noise is not eliminated and range of fitting is small.
Indeed a fitting of the data, obtained from original snapshots (not shown) to the form 
\begin{eqnarray}\label{fitting}
 \ell(t)=C+At^{\alpha},
\end{eqnarray}
in the range $[0,20000]$ MCS gives $\alpha=[0.15,0.25]$ (the value being larger when 
$\ell(t)$ is calculated from (\ref{lt_Crt}) or (\ref{lt_Skt})) which is consistent with 
earlier reports \cite{MRao,Grest}.
On the other hand a similar fitting to the data obtained after removing the noise 
gives $\alpha=[0.3,0.34]$ and within statistical deviation, does not depend upon the range of fitting. 
 This latter result is already suggestive of absence of strong correction to 
the scaling. However, since data fitting is always not a very reliable exercise as will be discussed later, 
to further substantiate the claim about small correction to scaling, we take the route 
of finite-size scaling analysis that will also be useful in quantifying the finite-size effect.
\subsection{Formulation of Finite-Size Scaling}\label{fs_analysis}
~In equilibrium critical phenomena, the singularity of a quantity $Z$ is characterized 
in terms of $\epsilon =|T-T_{c}|$, temperature deviation from the critical point, as
\begin{eqnarray}\label{Zeqn}
 Z \approx Z_0 \epsilon^{z} \approx Z'_0 \xi^{-z/\nu};~ Z'_0 = Z_0 \xi_{0}^{z/\nu}
\end{eqnarray}
where the correlation length $\xi$ grows as
\begin{eqnarray}\label{xi_eqn}
 \xi\approx \xi_0 \epsilon ^{-\nu},
\end{eqnarray}
with $z$ and $\nu$ being the critical exponents.
However, for finite values of $L$ any critical enhancement is restricted 
and $Z$ is smooth and analytic. Such finite-size effects may appear as a difficulty in 
understanding results from computer simulations.
 However, this problem can be tackled by writing down finite-size scaling ansatz, thus 
accounting for the size effect, as
\begin{eqnarray}\label{ansatz}
 Z\approx Y(x)\epsilon ^{z}=Y'(x)\xi^{-z/\nu}.
\end{eqnarray}
In Eq. (\ref{ansatz}), $Y(x)$ is the finite-size scaling function that depends upon the 
scaled variable $x=L/\xi$ and is independent of system size. Note that $Y$ and $Y'$ differ by
a factor originating from different amplitudes $Z_0$ and $Z'_0$ used in Eq. (\ref{Zeqn}).
 In further discussion, however, we will remove primes from both $Z_0$ and $Y$ and distinction can be derived from
whether the scaling forms are written in terms of $\epsilon$ or $\xi$. 
\par
~At this stage, it is important to ask about the behavior of $Y$ as a function of $x$.
While for static quantities, such question is already addressed, for dynamics, where the 
finite-size effects are found \cite{ Sengers} to be much stronger, there 
is no appropriate understanding of the variation of $Y(x)$. Nevertheless, one can write down 
the following limiting behaviors:
\begin{eqnarray}\label{limit}
 \mbox{for}~ x \rightarrow 0~(\epsilon \rightarrow0; L<\infty),~~
Y(x)\sim x^{-z/\nu},
\end{eqnarray}
such that $Z$ is finite at criticality :
\begin{eqnarray}\label{Zfinite}
 Z\sim L^{-z/\nu}.
\end{eqnarray}
Eq. (\ref{Zfinite}), when compared with Eq. (\ref{Zeqn}), is consistent with the fact 
that at criticality $\xi$ is the only important length in the problem and $L$ must scale with $\xi$.
Keeping this important fact in mind, in fact, one can write (\ref{ansatz}) as
\begin{eqnarray}\label{Zeqn2}
Z \approx Y(x)L^{-z/\nu}.
 \end{eqnarray}
On the other hand,
\begin{eqnarray}\label{limit2}
 \mbox{for}~ x \rightarrow \infty~(L\rightarrow \infty, \epsilon >0),~~
Y(x)=Z_0,
\end{eqnarray}
 so that Eq. (\ref{Zeqn}) is recovered in the thermodynamic limit.
\par
~With the knowledge of $\nu$, Eq. (\ref{Zfinite}) can be used to estimate $z$ by calculating
$Z$ at $T_{c}$ for various system sizes. A better strategy however is to study $Z$ 
at finite-size critical points $T_{c}^{L}$, such that
\begin{eqnarray}\label{ZTc}
 Z|_{T_{c}^{L}}\sim L^{-z/\nu}; ~~T_{c}^{L} -T_{c}\sim L^{-1/\nu},
\end{eqnarray}
though true meaning of a critical point can be assigned to $T_{c}^{L}$ only in the
limit $L\rightarrow \infty$.
\begin{figure}[htb]
\centering
\includegraphics*[width=0.38\textwidth]{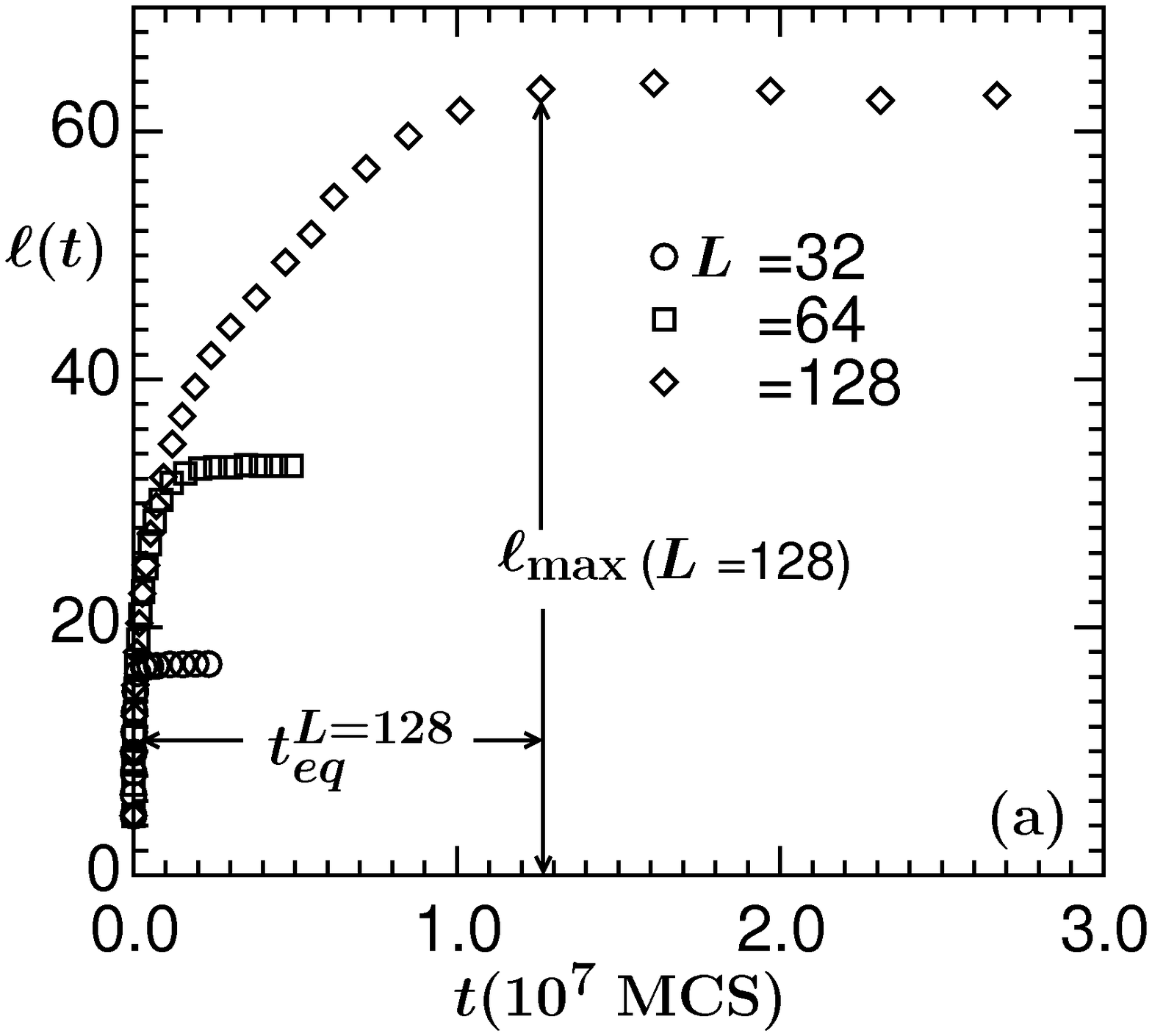}
\includegraphics*[width=0.37\textwidth]{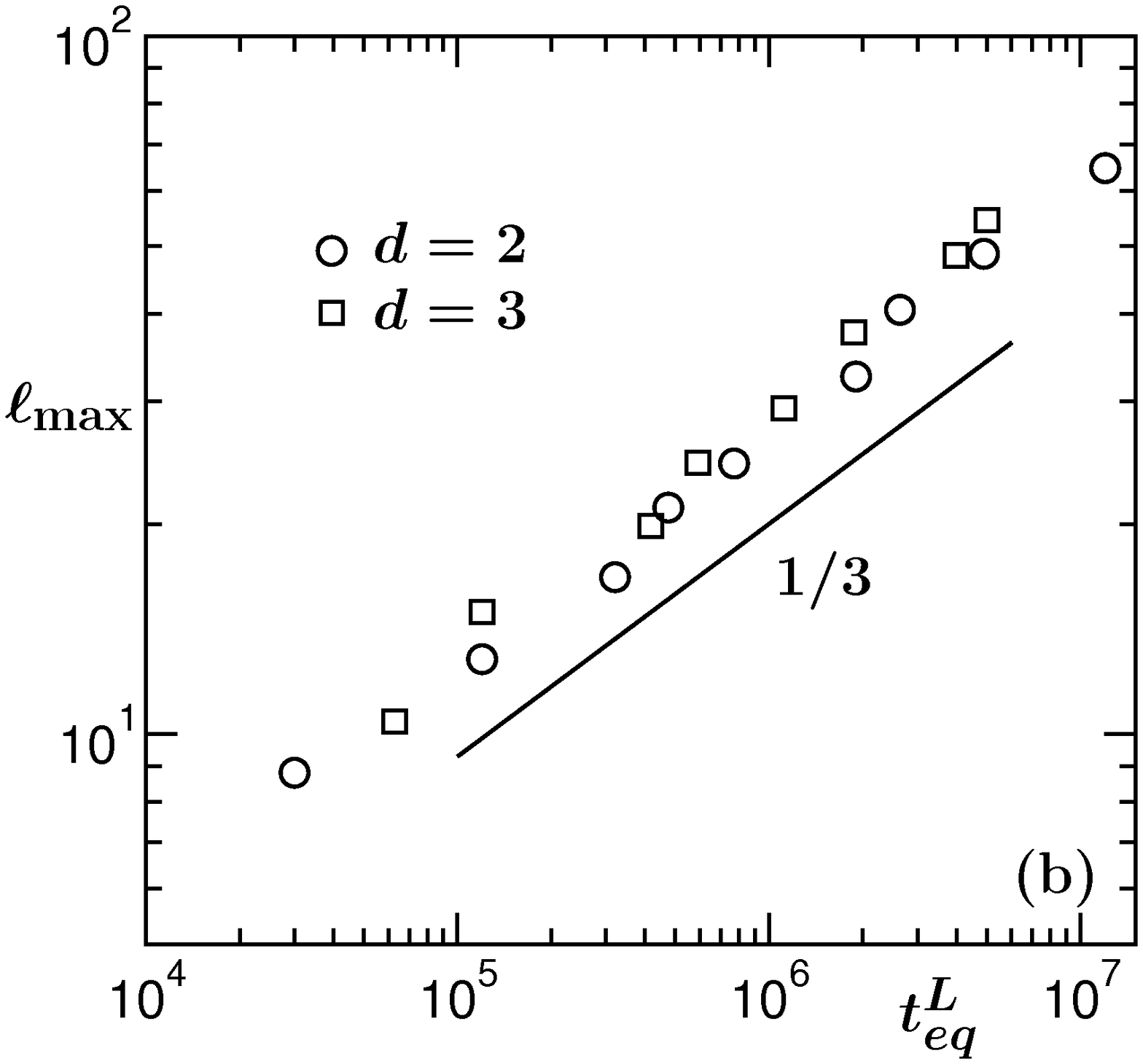}
\caption{\label{fig4}(a) Plot of average domain size $\ell(t)$, obtained from 
Eq. (\ref{L_d}), for the $2-d$ Ising model, for different system sizes 
(indicated on the figure) at $T=0.6T_{c}$. Definitions of $\ell_{\max}$
 and $t_{eq}^L$ are demonstrated. Data for $L=32$ and $L=64$
were averaged over 1000 independent initial configurations whereas only 
40 different initial realizations were used for $L=128$. Note that all subsequent 
results in this paper are obtained at the same temperature as this. (b) Demonstration
of the scaling behavior (\ref{teq}) in $d=2$ and $3$.}
\end{figure}
\par
~This general discussion about finite-size scaling method could be used to construct 
the similar formalism for non-equilibrium domain coarsening problem by identifying $\ell(t)$
with $\xi$ and $1/t$ with $\epsilon$. In the present problem, $L$ should scale with $\ell(t)$, 
more precisely $\ell_{\max} \sim \ell(t)$, where $\ell_{\max}(L)$ is the equilibrium domain size for a 
system of size $L$. In Fig. \ref{fig4}(a)  we show plots of $\ell(t)$ vs $t$ for various different values of 
$L$, in $d=2$. The flat regions in the plots at late times correspond to  $\ell_{\max}$.
\par
~At this stage, we would like to quantify the domain growth as
\begin{eqnarray}\label{lteqn}
\ell(t')=\ell_{0}+At'^{\alpha},
\end{eqnarray}
where $\ell_{0}$ is the average cluster size when the system becomes unstable
to fluctuations at time $t_0$ since the quench. Of course our measurement of time starts from there, 
i.e. $t'=t-t_0$. Note that we do not assign a meaning of domain size to this quantity and this should be 
treated in a manner similar to a background quantity in critical phenomena that appears from 
small length fluctuations whose temperature variation is usually neglected. 
Having said that, scaling part in Eq. (\ref{lteqn}) is only $At'^{\alpha}$. Of course, when 
$\ell(t')$ is significantly large, subtraction of the microscopic length $\ell_{0}$ does not
bring in noticeable difference. However, in computer simulations, where one deals with small systems,
presence of $\ell_{0}$ can bring in completely different conclusions. 
\par
~Eq. (\ref{lteqn}) is valid only in absence of any finite-size effect. For a finite 
$\ell_{\max}(L)$, analogous to (\ref{ansatz}), one can write down the scaling ansatz as 
\begin{eqnarray}\label{ltansatz}
 \ell(t')-\ell_{0}=Y(x)t'^{\alpha}
\end{eqnarray}
where now
\begin{eqnarray}\label{x}
 x=\frac{\ell_{\max}-\ell_{0}}{t'^{\alpha}}
\end{eqnarray}
is the scaling variable. Both in Eqs. (\ref{ltansatz}) and (\ref{x}), $\ell_{0}$ is 
subtracted to deal with the scaling parts only. By observing (\ref{Zeqn}), (\ref{limit}) 
and (\ref{limit2}) as well as (\ref{lteqn}), (\ref{ltansatz}) and (\ref{x}), one can arrive at 
the limiting forms of
$Y(x)$ as 
\begin{eqnarray}\label{Yx}
 Y(x)&\approx& x,~ \mbox{for}~ x \rightarrow 0(t'\rightarrow \infty, 
\ell_{\max} < \infty)
\end{eqnarray}
and
\begin{eqnarray}\label{Yx1}
Y(x)&=& A,~\mbox{for}~ x \rightarrow \infty (t' < \infty,~ \ell_{\max}\rightarrow\infty).
\end{eqnarray}
Of course, it would again be interesting to learn about the full form of $Y(x)$.
\par
~Also, analogous to $T_{c}^{L}$ in critical phenomena, one can define a finite 
size equilibration time $t_{eq}^{L}$ that is needed to reach $\ell_{\max}$, as 
demonstrated in Fig \ref{fig4}(a). Then one can write down a scaling equation analogous 
to (\ref{ZTc}) as
\begin{eqnarray}\label{teq}
[\ell_{\max}-\ell_{0}]\sim {t_{eq}^L}^{1/3}.
\end{eqnarray}
This scaling behavior is demonstrated in Fig. \ref{fig4}(b) where we show plots of $\ell_{\max}$ 
vs $t_{eq}^L$ on a log scale, in both $d=2$ and $d=3$. Consistency of the simulation data with 
the solid line of the form (\ref{teq}) confirms the validity of this approach. 
Note that in this figure, we did not subtract $\ell_{0}$ and the corresponding microscopic time
from the abscissa. As will be seen later, $\ell_{\max}$
 for the systems considered here are significantly larger than $\ell_{0}$, so one does not 
expect a big difference after subtracting. Eq. (\ref{teq}), of course, is a statement of the 
fact that dynamic critical exponent for this model is $3$.
\section{Simulation Results and Analysis}\label{results} 
~Having set the methodology in place, in this section we present results from MC simulation
of Kawasaki-Ising model in $d=2$ and $3$, combined with the finite-size scaling analysis.
\subsection{Results in \textit{d}=2}\label{2dresults}
\begin{figure}[htb]
\centering
\includegraphics*[width=0.375\textwidth]{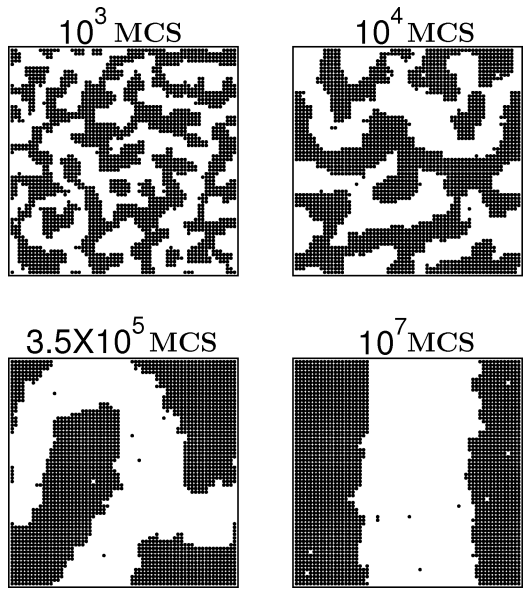}
\caption{\label{fig5} Evolution snapshots from different times, as indicated, 
for the Kawasaki-Ising model in $d=2$ at $T=0.6T_c$. The last snapshot corresponds to a completely 
equilibrated configuration.}
\end{figure}

~In Fig. \ref{fig5} we present snapshots during the evolution of an Ising system, starting from a 
$50:50$ random mixture of up and down spins, obtained via MC simulation at 
temperature $T=0.6T_{c}$. The times at which the shots were taken are mentioned on the figure. 
While the last snapshot corresponds to a situation when A and B phases are completely 
separated, the one at $t=3.5\times10^{5}$ MCS represents the situation when finite-size effect 
began to enter, which will be clear from subsequent discussion.
\par
\begin{figure}[htb]
\centering
\includegraphics*[width=0.35\textwidth]{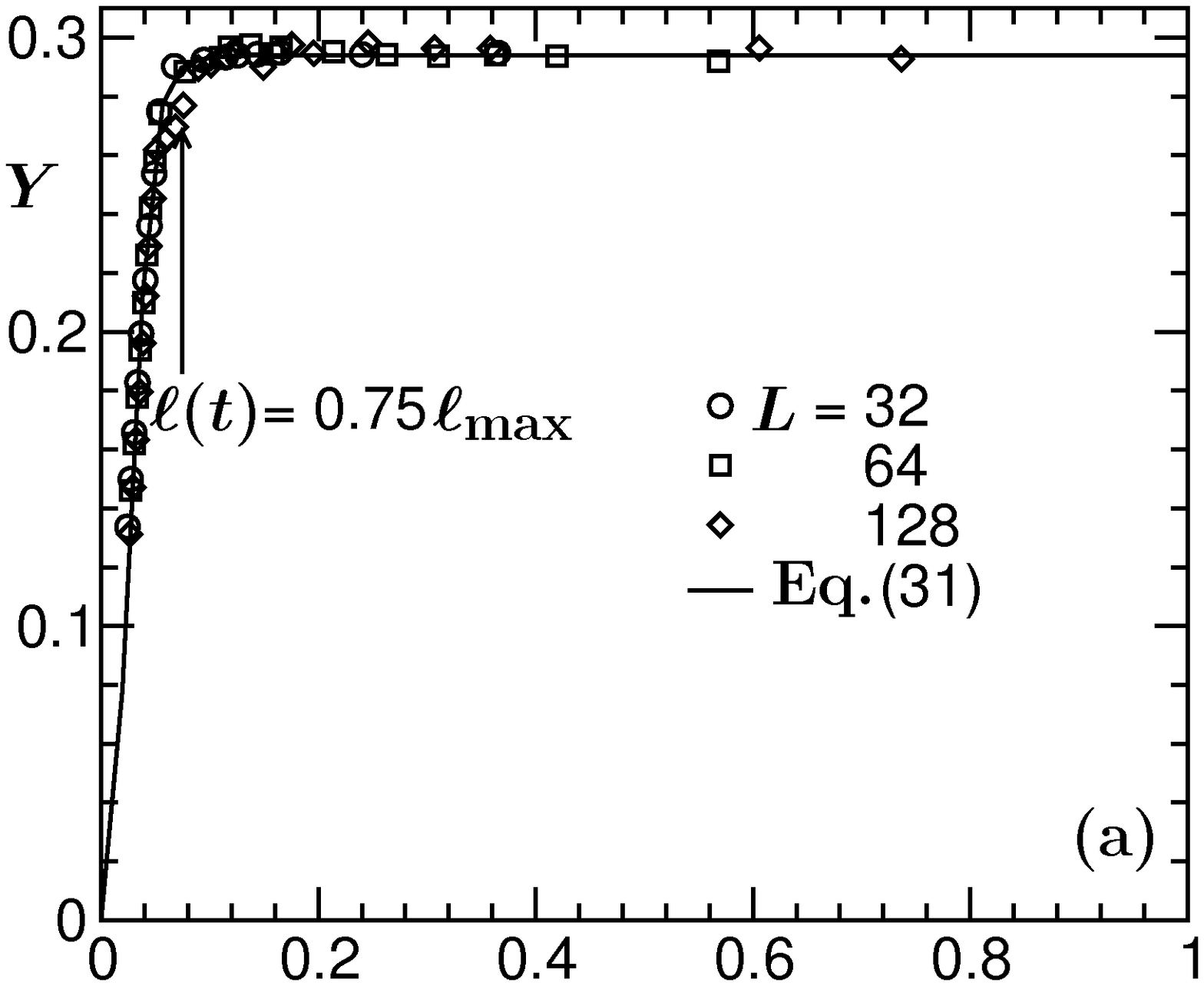}\\
\vskip 0.2 cm
\includegraphics*[width=0.35\textwidth]{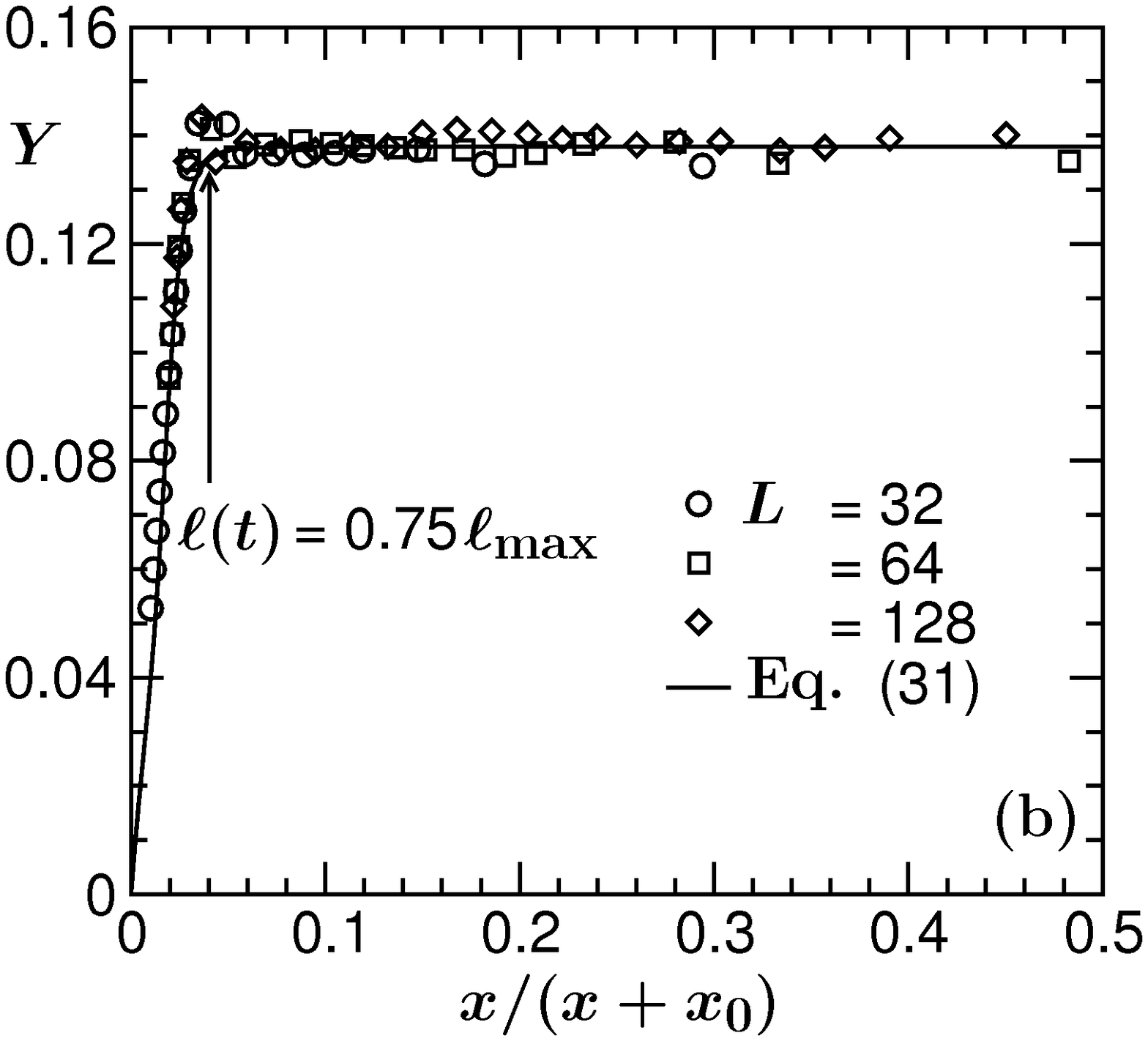}
\caption{\label{fig6} (a) Finite-size scaling plot of $Y$, with $\ell_{0}= 3.6$
lattice constants (after $20$ MCS from the quench time) and $\alpha\simeq 0.33$,
as a function of $x/(x+x_{0});~x_{0}=5$. The continuous curve is a fit to Eq. (\ref{Yfunc}) 
with the best fit parameters mentioned in the text. The arrow roughly marks the 
appearance of finite-size effect. Note that $\ell(t)$ data used here, were obtained 
from Eq. (\ref{L_d}). (b) Same as (a) but $\ell(t)$ obtained from the first zero-crossing 
of $C(r,t)$ [cf. Eq. (\ref{lt_Crt})]. In this case $\ell_{0} \simeq 2.7$ lattice constants (at $20$ MCS from quench)
and $\alpha \simeq 0.35$.}
\end{figure}

~From Fig. \ref{fig4}(a), it is already evident that finite-size effect is rather small. However, 
for a quantitative statement and to gain detail information about the growth exponent, 
more sophisticated analysis is called for. Following the discussion in the previous section, 
in Fig. \ref{fig6} we plot $Y=[\ell(t')-\ell_{0}]/t'^{\alpha}$ as a function of $x/(x+x_{0})$.
 Note that $x_{0}$ was introduced to see behavior of $Y$ properly both for small and large $x$. 
For convenience we set it to $5$. In this exercise we have varied $\alpha$ and $\ell_{0}$ 
(or the microscopic time $t_0$ associated with this length) to get optimum collapse 
of data from different system sizes. In Fig. \ref{fig6}(a), where $\ell(t)$
is being used from Eq. (\ref{L_d}), the optimum data collapse is obtained for 
$\ell_{0} \simeq 4a$ (average cluster size after $20$ MCS since quench), $a$ being the lattice spacing 
and $\alpha \simeq 0.33$. Similar exercise when $\ell(t)$ is being obtained from Eq. (\ref{lt_Crt}),
as shown in Fig. \ref{fig6}(b), gives $\alpha \simeq 0.35$ and $\ell_{0}$ there 
corresponds to the same number of MCS after quench. Note that $\ell_{0}$ in our analysis 
is a bare length independent of time and the scaling behavior (\ref{ltansatz}) will 
be obtained when this is chosen appropriately. These numbers, as expected, provided a 
constant value of $Y(x)$ in the region unaffected due to finite system size, which should 
be identified with the growth amplitude $A$ for which we quote $0.29\pm 0.01$ (cf. Fig. \ref{fig6}(a)). 
The arrows in Fig. \ref{fig6} marks the location where $Y(x)$ starts deviating from its constant value. 
The sharp nature of the crossover is indicative of only small size effect which we quantify from
the location of the arrow marks as
\begin{eqnarray}\label{lt_finite}
\ell(t)=(0.75\pm0.05)\ell_{\max}.
\end{eqnarray}
Of course, this value is significantly large compared to earlier understanding and expectation. Note that the 
snapshot at $t=3.5\times10^{5}$ MCS in Fig. \ref{fig5} corresponds to this length.
\par
~In an attempt to learn the full form of $Y(x)$, we construct the following 
functional form
\begin{eqnarray}\label{Yfunc}
 Y(x)=\frac{Ax}{x+1/(p+qx^{\beta})},
\end{eqnarray}
that has limiting behaviors consistent with (\ref{Yx}) and (\ref{Yx1}).
The continuous lines in Figs. \ref{fig6}(a) and (b) are fits to the form (\ref{Yfunc}) 
with 
\begin{eqnarray}\label{parameter1}
A\simeq0.29,~p \simeq 3,~q \simeq 6400,~\beta=4
\end{eqnarray}
 and 
\begin{eqnarray}\label{parameter2}
A\simeq0.14,~p \simeq 7,~q \simeq 13700,~\beta=4,
 \end{eqnarray}
thus have the convergence
\begin{eqnarray}\label{conv_eqn}
(x\rightarrow \infty)~Y(x)\approx A[1-fx^{-n}];n=5.
\end{eqnarray}
Of course, possibility of an exponential correction cannot be ruled out. This may be compared with much slower
convergence of such function in dynamic critical phenomena \cite{Sengers}. Note that the 
understanding of finite-size effect in both equilibrium and non-equilibrium dynamics 
is a non-trivial task and significant attention is called for.
\begin{figure}[htb]
\centering
\includegraphics*[width=0.35\textwidth]{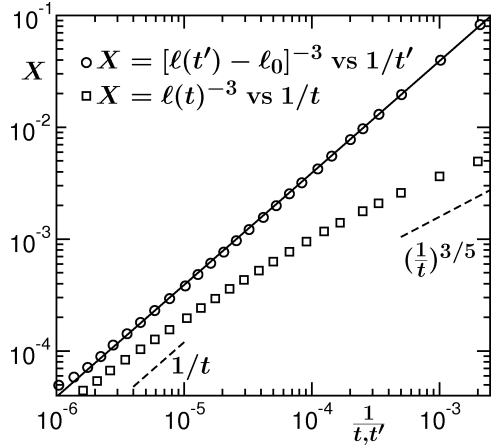}
\caption{\label{fig7} Plot of $[\ell(t')-\ell_0]^{-3}$ vs $1/t'$  and $\ell(t)^{-3}$ vs $1/t$ 
for $L^{2}=64^{2}$, with $\ell(t)$ being calculated from Eq. (\ref{L_d}). 
The continuous line has slope $39=1/A^3$.}
\end{figure}
\par
~To take a direct view of what happens after the corrective measure, in terms of 
subtraction of $\ell_{0}$, has been taken, in Fig. \ref{fig7} we plot
 $[\ell(t')-\ell_{0}]^{-3}$  vs $1/t'$ and $\ell(t)^{-3}$ vs $1/t$  
for $L=64$. A log-scale was used to bring visibility to a wide range of data. 
 The linear behavior of the data after 
subtracting $\ell_{0}$, starting from very early time justifies the introduction of 
$\ell_{0}$ again. The continuous line there is a plot of the form \textit{\~{A}}$x$ with 
\textit{\~{A}}$~\simeq 39=1/A^{3}$. On the other hand, notice the strong curvature when $\ell_{0}$
is not subtracted. The dashed lines marked by $1/t$ and $({\frac{1}{t}})^{3/5}$ on this figure corresponds
to $\ell(t) \sim t^{1/3}$ and $t^{1/5}$ respectively. Thus, when $\ell_0$ is not appropriately subtracted, only 
observing the trend on a log-log plot one may be misled to conclude that there is gradual crossover 
from one regime to the other. See Ref \cite{Puri_Oono} for a discussion on earlier belief about a crossover 
from $t^{1/4}$ to $t^{1/3}$.
Note that the exercise here as well as one in Fig. \ref{fig6}, where $Y$ is very flat from very
early time all the way to the moment when finite-size effect enters, are already 
indicative of absence of any strong corrections to scaling.
\begin{figure}[htb]
\centering
\includegraphics*[width=0.325\textwidth]{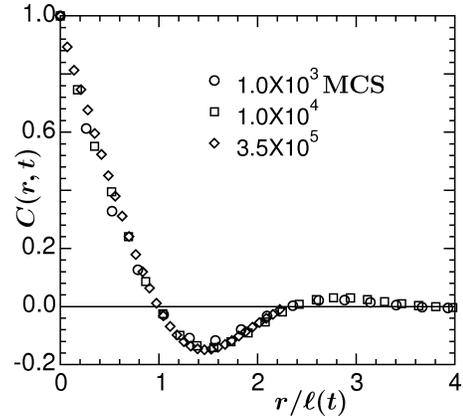}\\
\caption{\label{fig8} Scaling plot of $C(r,t)$ at $T=0.6T_c$. Note that $\ell(t)$
 was obtained using Eq. (\ref{L_d}).}
\end{figure}
\par
\begin{figure}[htb]
\centering
\includegraphics*[width=0.35\textwidth]{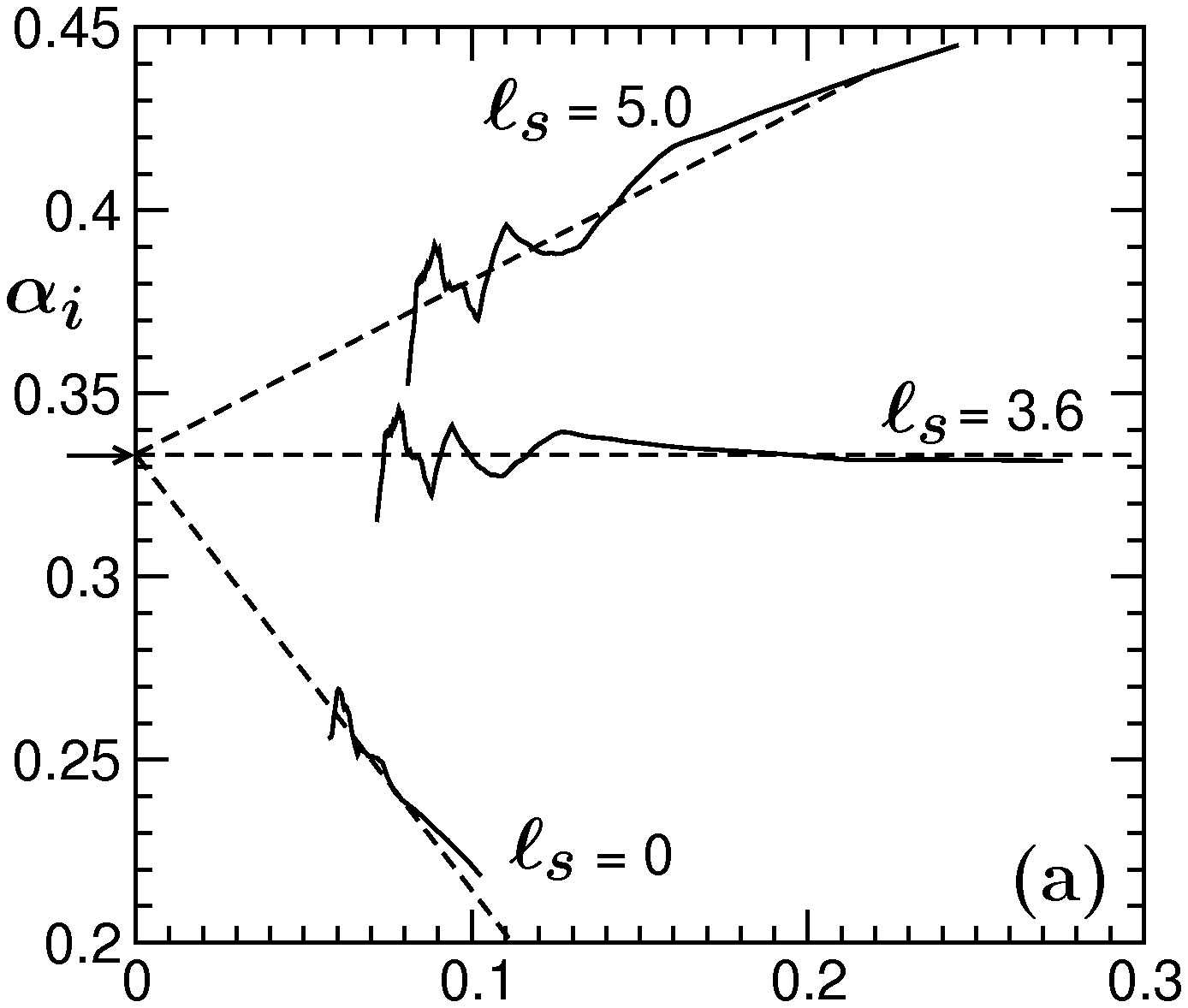}
\vskip 0.2cm
\includegraphics*[width=0.35\textwidth]{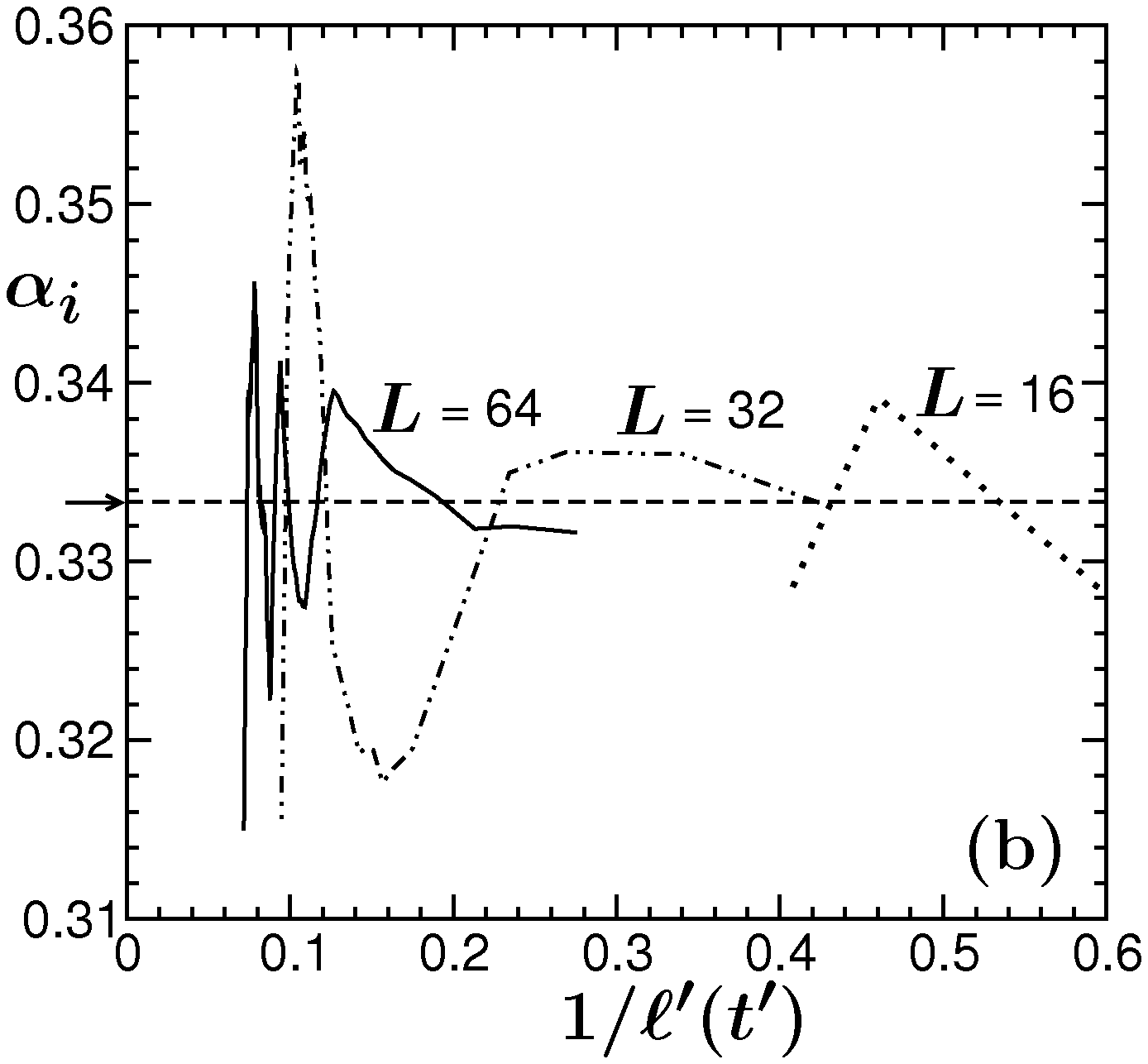}
\caption{\label{fig9} (a) Plot of instantaneous exponent $\alpha_{i}$ as a
function of $1/\ell'(t')$ for three different choices of $\ell_{s}$ as indicated, with $L^{2}=64^{2}$. 
The dashed straight lines have slopes $-1.19, 0$ and $0.49$, respectively. 
(b) Plot of $\alpha_{i}$ vs $1/\ell'(t')$ for $\ell_{s}=3.6$ and $L^{2}=16^{2}, 
32^{2}$ and $64^{2}$. In both (a) and (b) the arrows on the ordinate 
mark the value $\alpha=1/3$. Note that $\ell(t)$ was calculated from Eq. (\ref{L_d}).}
\end{figure}

~Before moving ahead for another elegant proof of the evidence for the absence of negligible 
corrections to scaling, we pass by showing the scaling plot of $C(r,t)$ in Fig. \ref{fig8} 
where good quality data collapse is obtained starting from very beginning 
till $t=3 \times 10^{5}$ MCS when the finite-size effect begins.
Next we introduce a length $\ell_{s}$ to write 

\begin{eqnarray}\label{ls}
  \ell'(t')=\ell(t')-\ell_{s}=[\ell_{0}-\ell_{s}]+At'^{\alpha},
 \end{eqnarray}
and calculate the instantaneous exponent \cite{Huse}
\begin{eqnarray}\label{alphai_def}
 \alpha_{i}=\frac{d[\ln \ell'(t')]}{d[\ln t']},
\end{eqnarray}
to obtain
\begin{eqnarray}\label{alpha_i}
 \alpha_{i}=\alpha\left[1-\frac{\ell_{0}-\ell_{s}}{\ell'(t')}\right].
\end{eqnarray}
According to Eq. (\ref{alpha_i}), when $\alpha_{i}$ is plotted as a function of 
$1/\ell'(t')$, for $\ell'(t')>0$, one expects linear behavior with a $y$-intercept equal to 
$\alpha$. Fig. \ref{fig9}(a) shows such plots for $\ell_{s}=0.0,3.6$, and $5.0$, as indicated. 
The dashed lines have $y$-intercept $\alpha=1/3$ and slopes 
\begin{eqnarray}\label{slopes}
 m=-\frac{\ell_{0}-\ell_{s}}{3}.
\end{eqnarray}
The consistency of the actual data with the dashed lines is remarkable, particularly  the behavior of 
$\alpha_{i}$ for $\ell_{s}=3.6$, again speaks for the choice of $\ell_{0}$ and strongly 
indicates that the LS scaling regime is realized very early. In Fig. \ref{fig9}(b) we 
present results with $\ell_{s}=3.6$ for various system sizes $L^{2}=16^{2}, 32^{2}$ and 
$64^{2}$. In all the cases, $\alpha_{i}$ oscillates around $1/3$. This observation, using 
a system size as small as $L^{2}=16^{2}$, stresses against unnecessary attempt to 
simulate larger systems. 
\par
This result is in strong disagreement with the earlier \cite{Huse} understanding of 
domain coarsening in $2-d$ conserved Ising model for critical quench that $\alpha$ is strongly time dependent 
and the LS value is recovered only asymptotically as 
$\ell(t\rightarrow \infty) \rightarrow \infty$.
The route to this finite-time correction was thought to be an additional term $\propto 1/\ell(t)^{3}$
in Eq. (\ref{dldt}) [cf. Eq. (\ref{Correction})], accounting for an enhanced interface conductivity. 
Note that a term $\propto 1/\ell(t)^{3}$ could also be motivated by introducing a curvature 
dependence in $\sigma$  as 
\begin{eqnarray}\label{sigma}
 \sigma[\ell(t)]=\frac{\sigma(\infty)}{1+\frac{2\delta}{\ell(t)}},
\end{eqnarray}
$\delta$ being the Tolman length \cite{Tolman}. However, our observation of negligible correction to the 
exponent, starting from the very early time, is consistent with the growing evidence \cite{Winter,Block} 
that Tolman length is absent in a symmetrical model \cite{Wortis} where the leading correction 
is of higher order. Also, small corrections that 
may be present, coming from the curvature dependence of the kinetic pre-factor in Eq. (\ref{dldt}) is 
beyond the accuracy of data in the present work. On the other hand, for $50:50$ composition, 
since the domain boundaries are essentially flat starting from very early time, any curvature dependence 
is expected to be absent. Thus, we conclude that this misunderstanding about the strong time dependence 
in $\alpha$ was due to the presence of a time independent length $\ell_{0}$ in $\ell(t)$ which 
our analysis subtracts out in appropriate way.
\begin{figure}[htb]
\centering
\includegraphics*[width=0.5\textwidth]{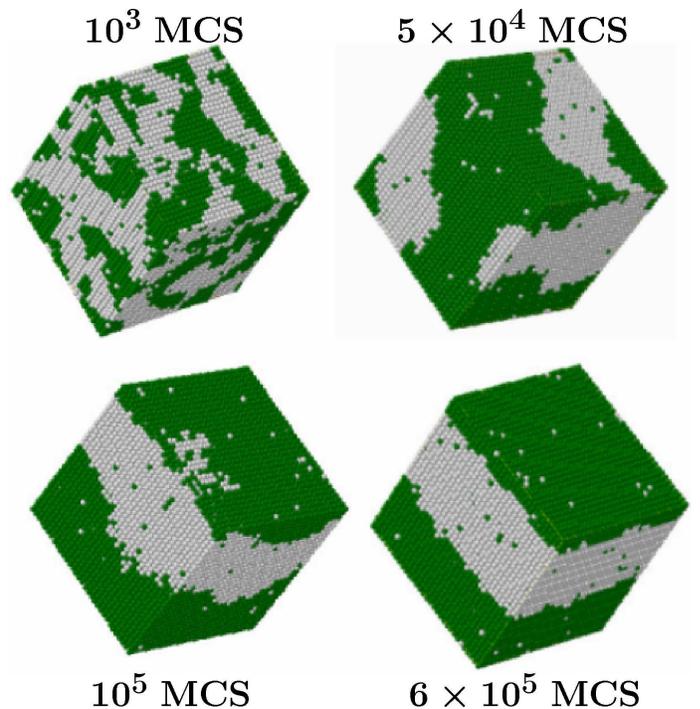}
\caption{\label{fig10} Evolution snapshots from different times for $3-d$ 
Ising model with $L^{3}=32^{3}$ and $T=0.6T_{c}$. A and B particles are marked 
black and grey respectively.}
\end{figure}
\subsection{Results in \textit{d}=3}\label{3dresults}
~In this subsection we turn our attention to the kinetics of phase separation in $d=3$. 
Fig. \ref{fig10} shows $3-d$ snapshots of the time evolution of Kawaski-Ising model at 
four different times as indicated on the figure where the last snapshot is clearly seen to 
have been equilibrated. Analogous to $d=2$, all results presented here were obtained at
$T=0.6T_{c}$, with $T_{c}=4.51 k_{B}T/J$ in this case, and the composition was chosen to be 
$50:50$ as well.
\begin{figure}[htb]
\centering
\includegraphics*[width=0.36\textwidth]{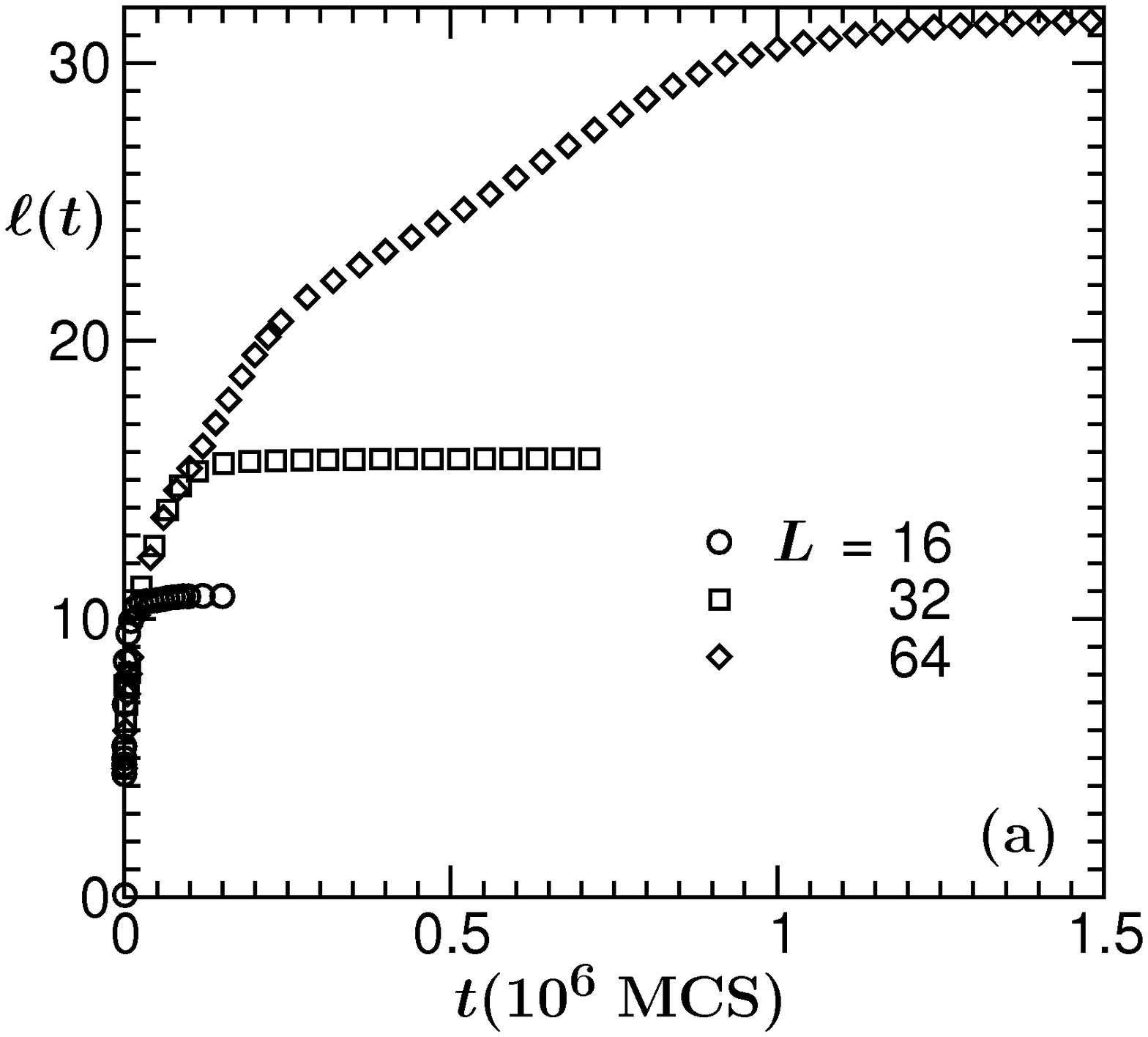}\\
\vskip 0.2cm
\includegraphics*[width=0.35\textwidth]{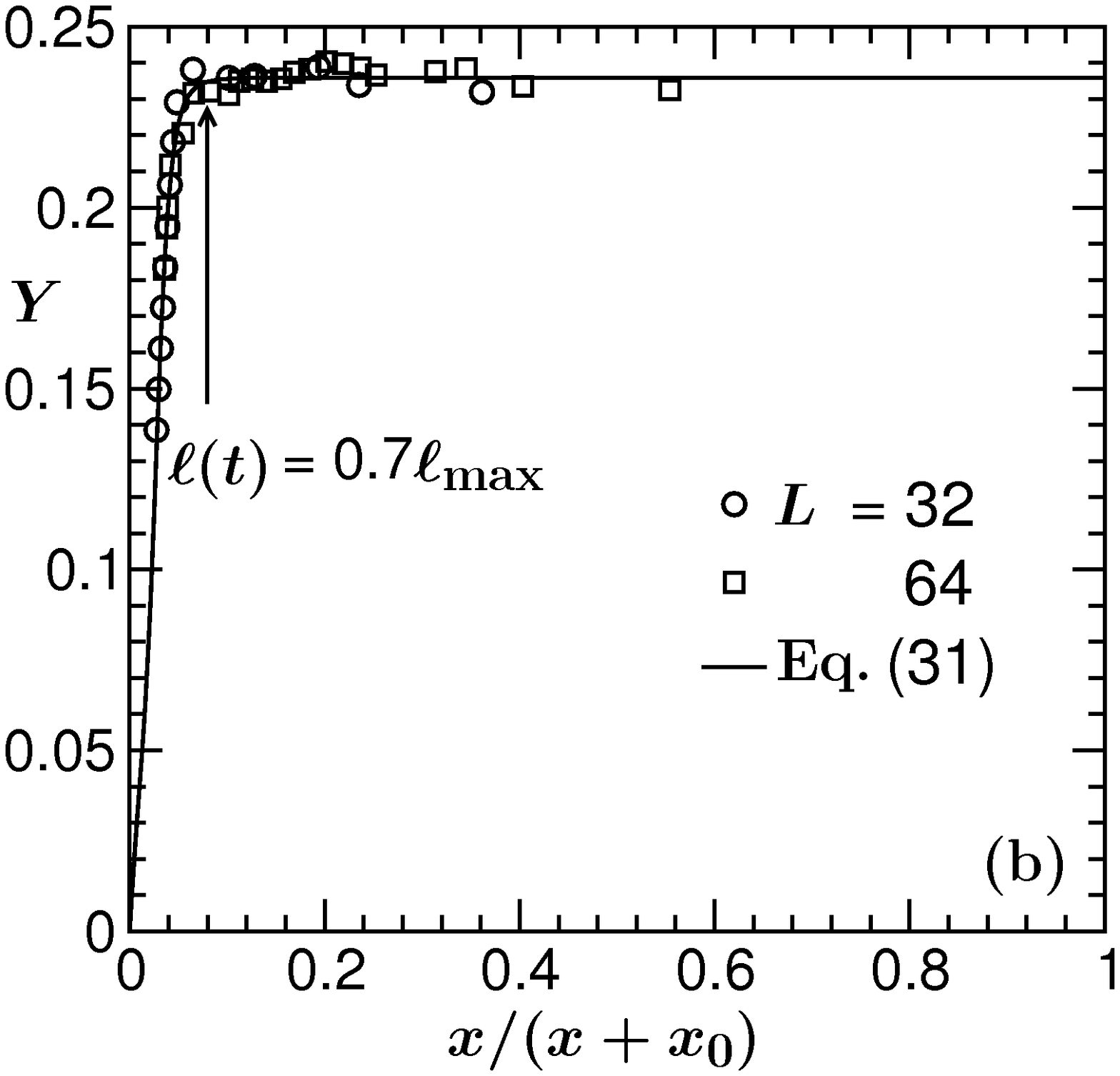}
\caption{\label{fig11} (a) Plot of $\ell(t)$, obtained from the first zero crossing 
of $C(r,t)$, vs $t$, for the systems $L^{3}=16^{3},32^{3}$ 
and $64^{3}$. (b) Finite-size scaling plot of $Y$, for the data presented in (a), vs 
$x/(x+x_{0})$ with $x_{0}=5$. Here $t_{0}=10$ ($10$ MCS from the quench) and $\alpha=0.35$. 
Appearance of finite size effect, obtained from the arrow mark, is estimated to be at 
$\ell(t)\simeq0.7 \ell_{\max}$, in close agreement with the one for $d=2$.}
\end{figure}
\par
~In Fig. \ref{fig11}(a) we present direct plots of $\ell(t)$ as a function of $t$, 
for $L^{3}=16^{3}, 32^{3}$ and $64^{3}$ where $\ell(t)$ was calculated from Eq. (\ref{lt_Crt}). 
Again, the finite-size effects look to be small. In Fig. \ref{fig11}(b) we present 
a plot of $Y(x)$, using the data in Fig. \ref{fig11}(a), as a function of 
$x/(x+x_{0});x_{0}=5$. Best data collapse in this case was obtained for 
$\ell_{0}=2.5$ ($10$ MCS after the quench, note that corresponding value of $\ell_{0}$
from Eq. (\ref{L_d}) is $3.0$ and $\alpha \simeq 0.315$) and $\alpha \simeq 0.35$. 
Very flat behavior of $Y(x)$, starting from 
the beginning again speaks for absence of any strong correction to the growth law. 
However, compared to the $d=2$ case, one may expect slightly stronger correction here
because of the inherent curvature present in the cylinder like domain objects as opposed to 
the stripe like structures in $d=2$. Possibly because of that  we could not obtain good collapse 
of data from $L^3=16^3$ on top of the ones presented, since the whole data set for $L^3=16^3$ 
is from very early time and suffers from corrections. Onset of finite size effect, as 
obtained from the arrow mark where $Y(x)$ deviates from the flat behavior, is in 
quantitative agreement with the $2$-dimensional situation, as quoted in 
Eq. (\ref{lt_finite}). Here also the third snapshot in Fig. \ref{fig10} 
(at $t=10^{5}$ MCS) is presented at this onset. A fitting, shown by the continuous line, to the form 
(\ref{Yfunc}) [$A=0.24$, $p \simeq 4$, $q \simeq 13050$ and $\beta=4$], again is consistent with asymptotic 
convergence (\ref{conv_eqn}).
\begin{figure}[htb]
\centering
\includegraphics*[width=0.35\textwidth]{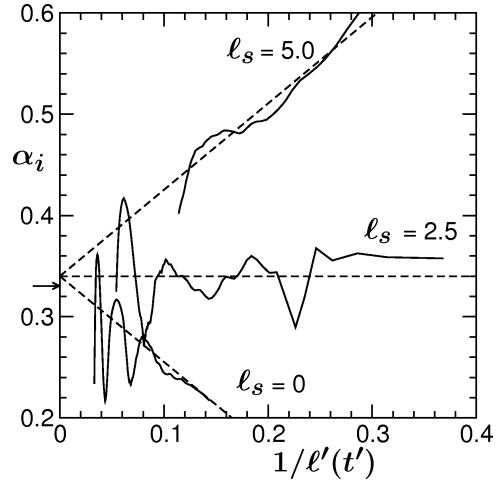}
\caption{\label{fig12} Plot of instantaneous exponent $\alpha_{i}$ vs $1/\ell'(t')$ 
with three different values of $\ell_{s}=0,2.5$ and $5$. The dashed lines correspond to $\alpha=0.34$.
The arrow on the ordinate marks the value $\alpha=1/3$.}
\end{figure}
\par
~ In Fig. \ref{fig12}, we present instantaneous exponent $\alpha_{i}$ as a function of $1/\ell'(t')$
for $L^3=64^3$ and three choices of $\ell_s$ as indicated. In all the cases, 
the exponent fluctuates around mean value $0.34$. Note that $\alpha$ estimated from 
$S(k,t)$ and $P(\ell_{d},t)$ are slightly higher and lower, respectively, compared 
to the one presented.
\par
~The appearance of growing oscillation in $\alpha_{i}$, seen in Figs. \ref{fig9} and 
\ref{fig12}, around the mean value was 
also pointed out by Shinozaki and Oono \cite{Shinozaki}. In a finite system, 
as time increases, for an extended period of time two large neighboring domains of same 
sign may not merge, thus lowering the value of $\alpha$. After a long time when 
two large domains merge, brings in drastic enhancement. This character is in fact visible 
in the direct plot of $\ell(t)$ vs $t$ at late times [cf. $L=128$ in Fig. \ref{fig4}(a)
 and $L=64$ in Fig. \ref{fig11}(a)]. 
Note that this oscillation could be a route to an error if one obtains $\alpha$ from least
 square fitting without choosing the range appropriately. 
Finally, it will be interesting to know the temperature dependence of $\ell_{0}$ and 
amplitude $A$ as well as of finite size effects. 
All these, however, we leave out for future work.
\section{Summary}\label{summ}
~This paper contains comprehensive study of domain coarsening in a 
phase separating system with diffusive dynamics in $d=2$ and $d=3$. Various 
different ways of analysis give results for growth law consistent with the expected LS exponent 
$\alpha=1/3$. As opposed to the earlier understanding, correction appears to be very weak, 
thus LS scaling behavior being realized very early. Small primary finite-size effect is a
welcome message which is suggestive of avoiding large systems, rather focusing on accessing
 long time scale which often is necessary for systems exhibiting multiple scaling regimes.
\par
~Our observation should be contrasted with an earlier study of Heermann, Yixue and Binder 
\cite{Heermann} that reports very strong finite-size effect. However, this latter study was based on 
an extremely off-critical composition and should not be considered to have general validity.
Note that due to the expected presence of correction in such off-critical composition, 
where droplet like structures form with finite radius of curvature at early time, the 
analysis is more difficult. Also, one should be prepared to encounter 
stronger size effect in more complicated situations, e.g., systems exhibiting 
anisotropic patterns \cite{Horbach,Mitchell,Bucior,Yelash,Yelash1,Hore,Das_Stat}. 
\par
~One may of course ask if the small finite-size effect observed for diffusive dynamics is 
also valid for kinetics of phase separation in fluids. A comprehensive study in that direction, 
for both binary and single-component fluids, is in progress. Such studies are important since 
brute force method of simulating very large systems, particularly for the study of fluid phase 
separation via MD simulation, is not often helpful to access long time scales even with the present 
day high speed computers and thus may not bring very conclusive understanding. 
\par
~A deeper understanding of $\ell_{0}$ requires further study, particularly, how the system is led 
to instability is a fundamental question to be asked. Even though scaling corrections appear to 
be negligible for critical quench due to the flat nature of the domain boundaries, one 
expects corrections for off-critical composition. This expected correction coming from 
surface tension should be of higher order than linear for a symmetric model. On 
the other hand, it would be interesting to learn about the leading order correction 
coming from the kinetic pre-factor.
\par
~Finally, we expect the observation, understanding and finite-size scaling technique used in this work to find relevance 
in other systems exhibiting growing length scales, e.g., ordering in ferromagnets, surface 
growth, clustering in cooling granular gas, dynamic heterogeneity in glasses, etc. 
In line of this work many earlier studies on domain coarsening may need to be 
revisited for better understanding which was not  possible because of lack 
of reliable methods of analysis.
\section*{Acknowledgment}\label{ack}
~SKD is grateful to K. Binder for useful discussion. The 
authors acknowledge financial support via grant number SR/S2/RJN-13/2009 of the 
Department of Science and Technology, India. SM also acknowledges Council of Scientific and Industrial Research, 
India, for financial support in the form of research fellowship. 
\par
${*}$ das@jncasr.ac.in

\end{document}